\documentclass[prl,twocolumn,preprintnumbers,amsmath,amssymb]{revtex4}

\usepackage{graphicx}
\usepackage{dcolumn}
\usepackage{bm}
\usepackage{color}
\usepackage{amsmath}

\usepackage{array}
\usepackage{epstopdf}

\begin{document}

\title{Porous Superhydrophobic Membranes: Hydrodynamic Anomaly in Oscillating Flows}

\author{S. Rajauria$^{1,2}$}
\author{O. Ozsun$^{3}$}
\author{J. Lawall$^{4}$}
\author{V. Yakhot$^{3}$}
\author{K. L. Ekinci$^{3}$}
\affiliation{$^{1}$Center for Nanoscale Science and Technology, National Institute of Standards and Technology, Gaithersburg, Maryland 20899, USA.} \affiliation{$^{2}$Maryland NanoCenter, University of Maryland, College Park, MD 20742, USA.} \affiliation{$^{3}$Department of Mechanical Engineering, Boston University, Boston, Massachusetts 02215, USA.} \affiliation{$^{4}$Atomic Physics Division, National Institute of Standards and Technology, Gaithersburg, Maryland 20899, USA.}

%
%
\email[e-mail:] {ekinci@bu.edu}

\begin{abstract}
We have fabricated and characterized a novel superhydrophobic system, a mesh-like porous superhydrophobic membrane with solid area
fraction $\Phi_s$, which can maintain intimate contact with outside air and water reservoirs simultaneously. Oscillatory hydrodynamic measurements on porous superhydrophobic membranes as a function of $\Phi_s$ reveal surprising effects. The hydrodynamic mass oscillating in-phase with the membranes stays constant for $0.9\le\Phi_s\le1$, but drops precipitously for $\Phi_s < 0.9$. The viscous friction shows a similar drop after a slow initial decrease proportional to $\Phi_s$. We attribute these effects to the percolation of a stable Knudsen layer of air at the interface.
\end{abstract}

\date{\today}
\maketitle



\maketitle

To completely describe the flow of a viscous fluid past a solid body, one must solve the Navier-Stokes equations inside the fluid subject to boundary conditions on the solid surface \cite{landau}. These boundary conditions cannot be obtained from hydrodynamics, but emerge from the microscopic interactions of fluid particles with  the surface. Consequently, they are not universal. It is well-established, for instance, that the commonly-assumed no-slip boundary condition can be violated \cite{laugaHANDBOOK} on both hydrophobic \cite{Vinogradova,Bouzigues,TyrrellPRL01,ZhangPRL07} and superhydrophobic   surfaces \cite{bizonneNATUREMATERIALS03, kimPRL06, bizonnePRL06, davisPHYSFLUID09}. The consequences of a shift in boundary condition from no-slip to partial-slip are vast. Many natural organisms  survive simply by virtue of  slip \cite{NEINHUIS, Bush,Gao}. Slip flows are expected to impact technology by enabling drag reduction in both laminar \cite{Bouzigues} and turbulent flows \cite{Rothstein09}. This list goes on.


\begin{figure}[htpb]
\begin{center}
\includegraphics[width=3 in]{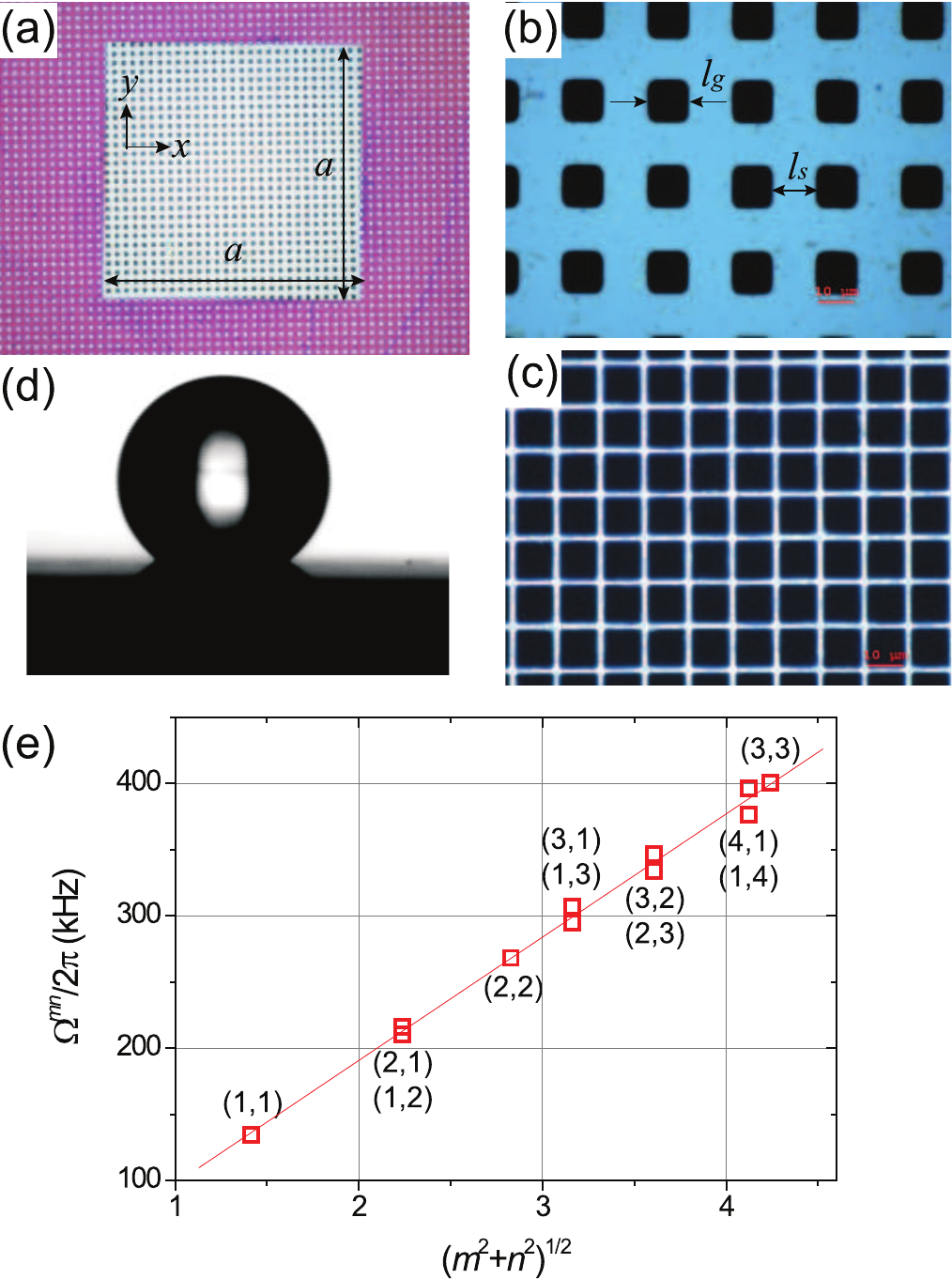}
\end{center}
\caption{(a) Top view of a porous membrane chip ($a\times a= 600\times600$ $\mu {{\rm{m}}^{\rm{2}}}$).
(b), (c) Optical micrographs of $\Phi_s=0.78$ and 0.34 membranes,
respectively. (d) A drop of water placed on a larger membrane
($a\times a=2\times2$ ${{\rm{mm}}^{\rm{2}}}$  and $\Phi_s=0.48$) showing the
superhydrophobicity of the surface. (e) Vacuum mechanical resonances of a $600\times600$ $\mu {{\rm{m}}^{\rm{2}}}$  porous membrane with $\Phi_s=0.34$.  Nearly-degenerate modes $(m,n)$ and $(n,m)$ are observed when $m\ne n$. Single standard deviations in the data are  smaller than the symbols.}
    \label{fig:1}
\end{figure}


On a conventional superhydrophobic surface \cite{degennesBOOK},  hydrophobicity combined with microscopic   roughness causes the water surface to remain suspended above the solid tips, with mostly trapped air underneath \cite{watanableLANGMUIR,bicoEPL01}. Since the flow is on a composite surface made up of solid and air, one solves the Navier-Stokes equations subject to  no-slip on the solid elements and to slip  at the water-air interface \cite{laugaHANDBOOK}. Thus, in a first pass analysis, viscous friction force on a superhydrophobic surface is found to be proportional to the wet solid area, $\Phi_{s}$ \cite{ybertPHYSFLUID04}.  In this manuscript, we show that flow on a porous superhydrophobic membrane deviates from the above picture. Oscillatory hydrodynamic response \cite{ekinciLoC10} of  the membrane suggests that a stable Knudsen layer of gas percolates on the membrane, changing the boundary condition. This is because the porous superhydrophobic membrane structure enables the surrounding air to move ballistically to the interface with little resistance --- in contrast to a conventional superhydrophobic surface, where  trapped gas pockets are diffusively connected to a gas reservoir through macroscopic distances.

The novel system under study shown in Fig. 1 is a tension-dominated porous silicon nitride membrane made hydrophobic by silanization. The membrane has a (nominal) macroscopic area of $a\times a= 600\times600$ $\mu {{\rm{m}}^{\rm{2}}}$ and a nanoscale thickness of $t_s=200$ nm. A matrix of identical square pores with dimensions $l_g\times l_g = 10\times10$ $\mu {{\rm{m}}^{\rm{2}}}$  are lithographically etched in the membrane. The pitch is $l_g+l_s$, where $l_s$ is the width of the solid strips in between the pores, as shown in Fig. 1(b). This results in a solid area fraction $\Phi_s=1-\frac {l_g^2} {(l_g+l_s)^2}$. When a drop of water is placed on the porous membrane, it is supported by a composite surface of solid and gas (air); thus, wetting is not favored as shown in Fig. 1(d).

We first characterize the \emph{intrinsic} mechanical properties of the porous membranes. In order to eliminate any fluidic effects, we perform these measurements under vacuum. For a tension dominated square membrane ($a\times a\times t_s$), the frequency of the normal mode $(m,n)$ in vacuum is given by ${{\Omega _v^{mn}} \over {2\pi}} = {[{{{\sigma _s}{\pi ^2}} \over {{\Phi _s}{\rho _s}{t_s}{a^2}}}({m^2} + {n^2})]^{1/2}}$ \cite{Timoshenko}. Here, $\sigma_s$ is the tension, $\rho_s$ is the density, and $m$ and $n$ are two integers. The \emph{in vacuo} mode frequencies $\Omega _v^{mn} \over 2\pi$ of a $\Phi_s=0.34$ membrane are shown in Fig. 1(e). Here, the resonances are excited by a piezoelectric-shaker and detected  using a Michelson interferometer at a pressure of $10^{-2}$ Pa. The data  confirm that the tension-dominated membrane approximation holds well, even for a membrane with a very small solid fraction.  Given that the modes are well-separated in frequency, each mode $(m,n)$ can be modeled as a damped harmonic oscillator with effective mass $ M_{s} = \frac{\rho_s \Phi_{s} t_s a^{2}}{4}$ and stiffness $K_{s} = \frac{\sigma_{s} \pi^{2}}{4}(m^{2}+n^{2})$. Relevant mechanical parameters for the fundamental modes of all our membranes are displayed in Table 1.


\begin{table}
\caption{\label{tab:devices} Mechanical properties for the fundamental mode of the porous membranes.}
\begin{ruledtabular}
\begin{tabular}{cccc}

$\Phi_{s}$ & $\Omega^{11}/2\pi$ & $K_{s}$ & $M_{s}$ \\
  & (kHz) &  (N/m) & ($10^{-12}$ kg) \\
\hline
1 & 235 & 134 & 61.2\\
0.96 & 233 & 126 & 58.7\\
0.88 & 223 & 104.7 & 53.8\\
0.82 & 208 & 67.6 & 50.2\\
0.78 & 142 & 38.1 & 47.7\\
0.65 & 159 & 39.2 & 39.8\\
0.48 & 106 & 12.9 & 29.4\\
0.34 & 134 & 14.2 & 21.1\\
\end{tabular}
\end{ruledtabular}
\end{table}


\begin{figure*}[htpb]
\begin{center}
\includegraphics[width=7 in]{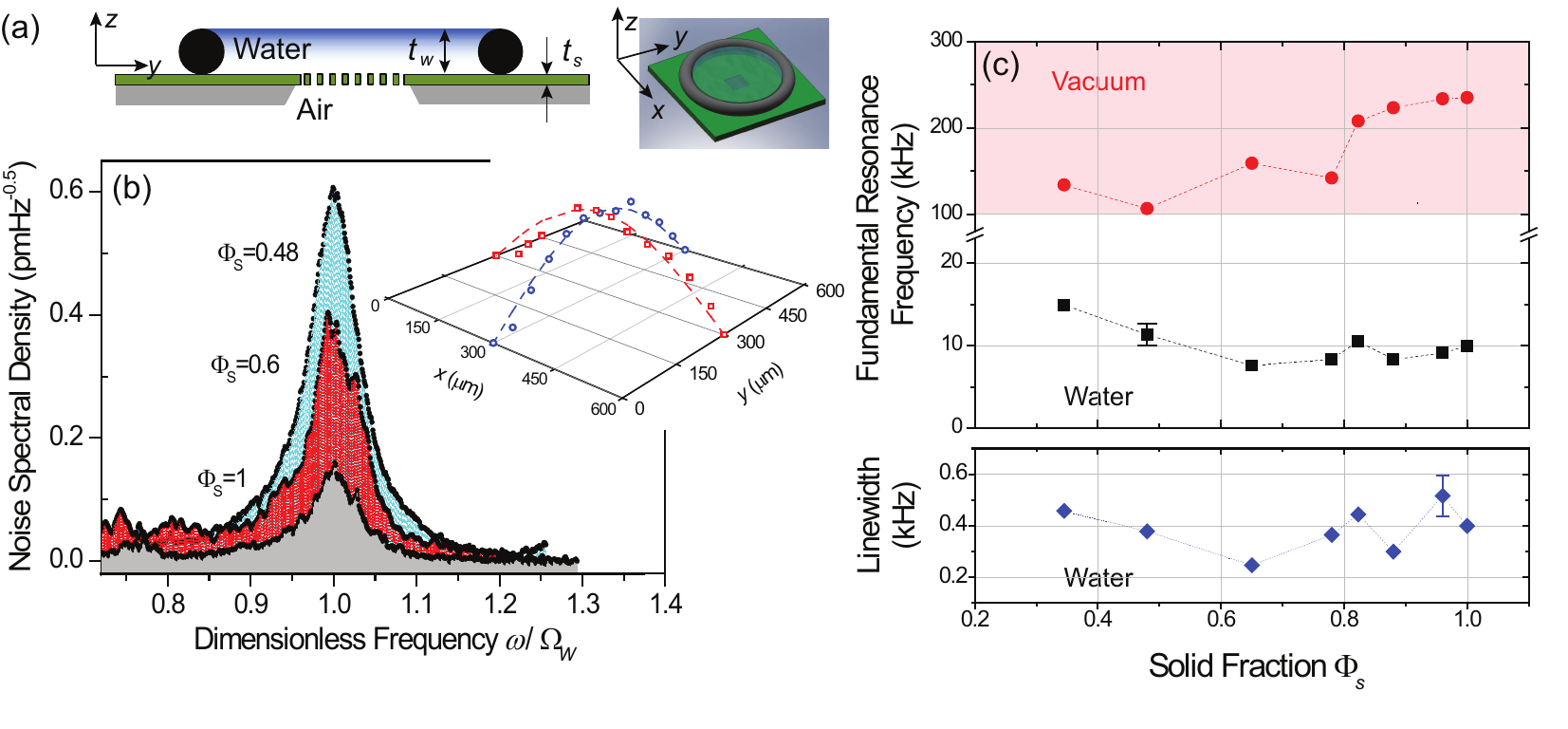}
\end{center}
\caption {(a) Schematic of the measurement cell in cross-sectional and isometric views. The cell is filled with
water and is placed on top of the membrane chip. A heterodyne Michelson interferometer probes the motion from below. (b) Thermal noise spectra of the fundamental mode of three different membranes with water atop. From top to bottom, $\Phi_s=$0.48, 0.6 and 1. The
frequency axis is normalized with the respective resonance frequencies in water.  The inset shows the shape of the fundamental mode for the $\Phi_s=0.6$ membrane. (c) Fundamental-mode resonance frequencies  in vacuum and with water atop, and linewidths $\gamma_w/2\pi$ with water atop. Error bars represent the associated single standard deviations and are only shown  when  larger than the symbols.}
    \label{fig:1}
\end{figure*}


We now turn to measurements with water. The measurements are performed using a fluid cell \emph{atop} the membrane as shown in
Fig. 2(a). The porous membrane does \emph{not} leak but  supports intimate and continuous contact with both the water reservoir above and the
gas reservoir (ambient atmosphere)  below. Using a heterodyne Michelson interferometer (with displacement sensitivity of 100 ${\rm{fm}}/\sqrt {{\rm{Hz}}}$ around 10 kHz  with 85 $\mu$W incident on the photodetector), we have measured the thermal-noise spectra of all the membranes in their fundamental modes. Fig. 2(b) shows the noise spectra measured at the center of three membranes with different $\Phi_{s}$. In order to confirm that we are working with the fundamental mode [$(m,n)=(1,1)$], we have scanned the optical spot along the $x$ and $y$ directions and obtained mode shapes, such as the one shown in the inset of Fig. 2(b). Since we exclusively study the hydrodynamic response of the fundamental mode here, we henceforth drop the superscript 11.  The top data trace in Fig. 2(c) shows all the fundamental resonance frequencies  in vacuum, ${\Omega_v}\over {2\pi}$, obtained by driving the membranes linearly. Thermal spectra with water atop the membranes have provided the resonance frequencies ${\Omega_w}\over {2\pi}$ and linewidths $\gamma_w\over {2\pi}$ (bottom trace) {\cite{Air}.


We first provide a general discussion of the fluid dynamics encountered in our system. We consider the out-of-plane  (broadside) oscillations of a rigid square ($a\times a$) plate \emph{immersed} in a viscous fluid. We take the plate velocity as the real part of the complex exponential, ${{\bf{u}}_{\bf{s}}} = \Re \left\{ {{U_s}{e^{i\omega t}}{\bf{\hat z}}} \right\}$, with amplitude $U_s$. Adopting the no-slip boundary condition, we find the magnitude of the fluidic force $F_f{\bf{\hat z}}$  on the plate in the high-frequency limit as \cite{ZhangStone}

\begin{equation}
{F_f} \approx \Re \left\{ {A{{{\mu _f}{a^2}} \over \delta }{u_s}{e^{i \phi }} + M_f{{d{u_s}} \over {dt}}} \right\},
\end{equation}

\noindent with $\phi\approx {{\pi} \over {4}}$ and $A\sim20$. The viscous boundary layer thickness, $\delta=\sqrt{\frac{2\mu_f}{\rho_f\omega}}$, depends on the dynamic viscosity $\mu_f$  and density $\rho_f$ of the fluid. $M_f$ is the so-called added or hydrodynamic mass, well-known from  the potential flow theory  around an accelerating solid body. Consequences of Equation (1) are as follows. Viscous energy dissipation is due to  tangential flow on the plate, expressed by the first term on the right-hand side. Being proportional to $\mu_f$, the dissipation provides a widely-used probe of the fluid-solid interaction. The second term on the right-hand side does \emph{not} contribute to dissipation since $u_{s}{{d{u_s}} \over {dt}}$ integrated over a cycle is zero. However, this term provides an independent probe of the fluid properties (near the solid) through $\rho_f$. To emphasize this, we write   $M_f=\rho_{f}V_{f}$, where $V_{f}$ stands for the volume of fluid displaced by plate motion and depends \emph{only} upon geometry. Indeed, it will be shown below that, in our system, the changes in the nature of the fluid near the solid boundary results in changes in both $M_{f}$ and dissipation.

Returning to the membrane oscillations, we make a one-dimensional harmonic oscillator approximation for the fundamental mode. We analyze all our experimental data (of Fig. 2 and Table 1)  using this approximation, obtaining the results shown in Fig. 3. In this approximation, the membrane has position $z_s$, velocity ${{\bf{u}}_{\bf{s}}} = {\dot z_s}{\bf{\hat z}}$, mass $M_s$ and stiffness $K_s$. We assume that $u_s$ is nearly sinusoidal because all membrane resonances in water have quality factors $Q_w\gtrsim 20$ and the thermal drive has a white spectrum: $u_s \approx {\Re}\left\{{U_s{e^{i\Omega_w t}}} \right\}$. Given that the dissipation from water dominates the overall dissipation \cite{Air}, we write ${M_s}\dot u_s \approx F_e+F_w$, where $F_e=-K_s z_s$ is the elastic spring force and $F_w=F_f$ in Eq. (1) with the appropriate parameters. Based on these considerations, we write a complex linear response function for the system as \cite{paul} $G(\omega ) \approx {[{K_s} - ({M_s} + {M_w}){\omega ^2} + i\alpha \omega ]^{ - 1}}$. The effect of the fluid is embedded in two measurable parameters: the added water mass $M_w$ and the friction coefficient $\alpha$ \cite{reif}.


The added water mass $M_w$ can be determined from the frequency shift of the membrane mode when it is loaded with water. The stiffness $K_s$ of the mode does not change appreciably from vacuum to water. Thus, ${M_s}{\Omega _v}^2 = ({M_s} + {M_w}){\Omega _w}^2$, which simplifies to ${M_s}{\Omega _v}^2 \approx {M_w}{\Omega _w}^2$ since ${M_s} \ll M_w$  \cite{maali}. Figure 3(a) shows $M_w$ as a function of $\Phi_s$, calculated using $M_s$ values in Table 1 and frequency values in Fig. 2(c). Note the two separate regions in Fig. 3(a) with a transition around $\Phi_s\approx 0.9$. In order to estimate $M_w$ from \textcolor[rgb]{1.00,0.00,0.00}{first principles}, we emphasize
that our resonator  is {\it not } immersed  in water [see Fig. 2(a)] --- unlike in a
typical set up. There is a water layer of thickness $t_w=2.4$ mm and density $\rho_w$
atop the membrane, but the backside is exposed to atmosphere. The
dominant hydrodynamic mass contribution comes from the the entire water
layer moving in-phase with the membrane in the $z$ direction \cite{shear-mass}. This provides
$M_w\approx {{\rho_{w}t_{w}a^{2}}  \over {4}} \approx 2\times
10^{-7}$ kg [dotted segment in Fig. 3(a)]. This estimate is in agreement with the data of Fig. 3(a), but \emph{only}
in the region $0.9\le\Phi_s\le1$. Further support for our estimate comes when
$t_w$ is reduced to approximately 1.2 mm, which results in a factor of 1/2 reduction in $M_w$. Experimentally, the  measured frequency increases by a factor of 1.2, which is close to the factor $\sqrt 2$ expected. For $\Phi_s \lesssim 0.9$, there is a significant deviation from this simple model: the measured $M_w$ shows a rather fast decrease, eventually by a factor of 23.


\begin{figure*}[htpb]
\begin{center}
\includegraphics[width=7 in]{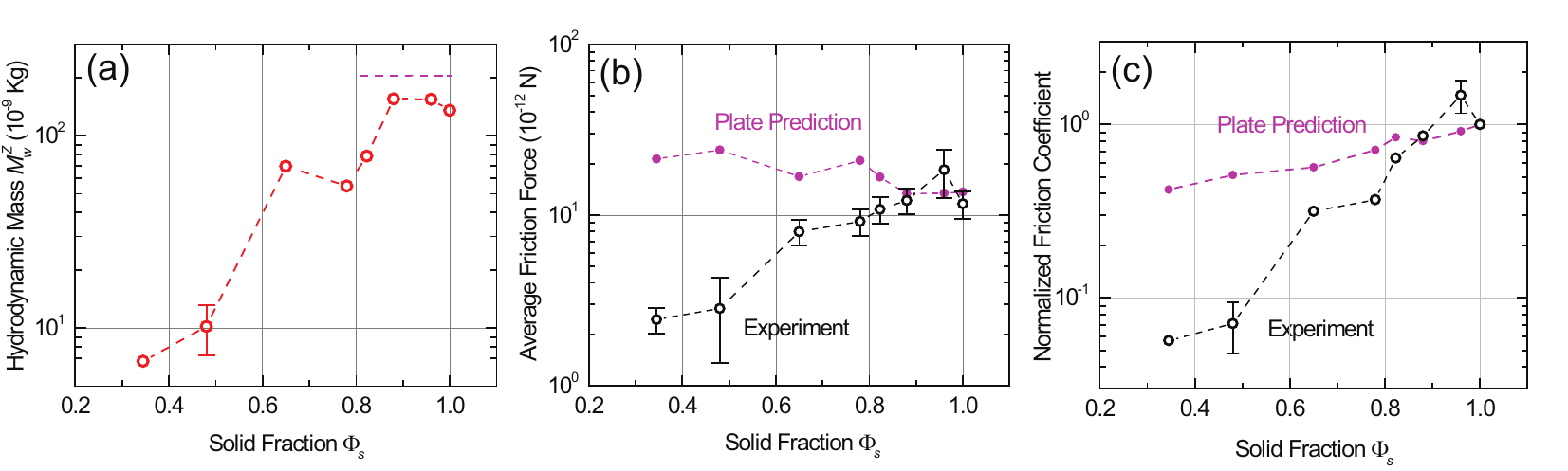}
\end{center}
\caption{(a) Measured $M_w$ as a function of $\Phi_s$. The line segment is the hydrodynamic mass of the entire water layer. (b) Average friction force and (c) the normalized friction coefficient. The plate prediction is calculated from Eq. (1)  using
experimental  velocities and frequencies where needed. The normalized friction coefficient for the plate model is $\approx \Phi_s$. Error bars represent the associated single standard deviations and are only shown  when larger than the symbols.}
    \label{fig:4}
\end{figure*}


The upper, slowly increasing trace in Fig. 3(b) is the
cycle-averaged viscous force  on the membrane, ${{A{\mu _w}{a^2}{\Phi _s}{U_s}} \over {2 {\pi
^2}\delta }}$, \emph{predicted} using the  plate model of Eq. (1) and accounting for the membrane mode-shape.
Here, $\mu_w$ is the dynamic viscosity of water;  ${A \over {2 {\pi ^2}}} \sim 1$; ${U_s}^2
\approx 2{\pi^2}{\Omega _w}^2\left\langle {{{z_s}^2}} \right\rangle $,
 ${\left\langle {{z_s}^2} \right\rangle ^{1/2}}$ being the thermal
amplitude of the membrane found from the integral of the measured
displacement noise  spectral density. It is important to note that, as
$\Phi_s$ becomes smaller, $U_s$ increases. This is because the
stiffness $K_s$ decreases (see Table 1), the average thermal
drive force remains constant and $\Omega_w$ changes very slowly. In the calculated plate model, the
decrease in the wet area, $a^2\Phi_s$, appears to be off-set by this
increase in $U_s$, thus resulting in a net increase in the drag
force as $\Phi_s$ decreases. The experimental cycle-averaged
friction force is obtained from the one-dimensional damped harmonic oscillator model as $\alpha U_s\approx M_w \gamma_w U_s$. This force plotted as the lower trace in Fig. 3(b) shows a surprising deviation
from the plate model. As in added mass, the
plate prediction agrees with the experiments when $\Phi_s\approx1$.
However, for  $\Phi_s\lesssim 0.9$, the drag force decreases rapidly,
attaining a value  an order of magnitude smaller than the plate
prediction at $\Phi_s=0.34$.

The drag reduction on the porous membranes can be better
assessed, if one considers the drag force per unit velocity: this is
the friction coefficient $\alpha $. Figure 3(c) shows
the predicted and experimentally-obtained normalized friction
coefficients, ${{\alpha ({\Phi _s})} \over {\alpha (1)}}$. The
predicted value is proportional to $\Phi_s$ since the system behaves
as a plate, but with a reduced solid area. The experimental values
are given by $\alpha \approx M_w\gamma_w$. The data show that drag
force for a given velocity can be reduced by a factor of 18, if one
goes from a complete membrane ($\Phi_s=1$) to $\Phi_s=0.34$,
i.e., $\alpha(0.34)\approx \alpha (1)/18$.


Given that the the membranes do not
leak and $M_w$ is constant for $0.9\le {\Phi _s} \le 1$, we conclude that  the presence of the air reservoir does not affect the flow
in this interval. The agrement between the predicted and measured friction
forces in the same interval provides more support for this conclusion. The significant deviation in the
measured response from the plate model for $\Phi_s\lesssim
0.9$ suggests that the flow changes around $\Phi_s\approx
0.9$. The new feature of our system is its openness to air at
atmospheric pressure. The membrane thickness, $t_s=200$ nm, is
close to the mean-free-path of air, $\lambda_{g}\approx 60$
nm. This enables  the surrounding air  to move through the membrane
pores with little resistance. The dramatic decrease in
$M_w$ for $\Phi_s<0.9$ can be attributed to a percolation
transition: air bubbles localized  within the well-defined pores
begin to coalesce as  $\Phi_s$ is decreased, eventually resulting in a complete gas layer, which
separates the solid strips from the water surface. This gas layer is expected to exist in the Knudsen regime, with its thickness $\xi$ smaller than its mean-free-path, $\xi\lesssim \lambda_{g}$.


Assuming a complete Knudsen layer at the interface, we can assess the friction reduction on a porous membrane with small $\Phi_s$.  A one-dimensional model will suffice. We consider a large porous plate  oscillating in its plane with velocity,  $\Re \left\{ {{U_s}{e^{i\Omega_w t}}{\bf{\hat x}}} \right\}$, under water with a Knudsen air layer in between the plate and the water. The velocity field inside the water is ${\Re}\left\{ {U_w{e^{- {z \over \delta }+i(\Omega_w t - {z \over \delta}) )}}{\bf{\hat x}}} \right\}$ and $U_{w}\ne U_s$. Since the stress is a continuous  function of coordinate at the interface ($z\approx0$), ${{{\mu _w}{U_w}}\over \delta } \sim {{{\rho _g}{U_s}{\Phi_s}{u_{th}}}\over 6}$.
Here, $U_s\Phi_s$ and $u_{th}$ are respectively the average hydrodynamic velocity and the thermal velocity of air molecules; $\rho_g$ is the density of air. The $1/6$ factor  accounts for the fraction of molecules traveling in the $+z$ direction. Using the parameters available, we derive $\frac{U_{s}}{U_{w}}\sim \frac{3}{\Phi_s}$.  The slip length \cite{laugaHANDBOOK}, ${\lambda} \sim {{{\mu _w}} \over {{\rho _g}{u_{th}}{\Phi _s}}}$, emerges as 6 $\mu$m at $\Phi_s=0.34$.

Our results might be relevant to applications. Unlike air bubbles on a hydrophobic surface \cite{BrennerPRL08}, the air layer in our system is stable against diffusion into the water because of the resistance-free influx from the air reservoir.  Assuming that porous pipes of macroscopic dimensions can be manufactured, significant drag reduction could be achieved.  Several puzzling phenomena in
bio-fluid-dynamics, including  transport through and over bio-membranes, and propulsion over the water surface, may
be related to the physics observed here \cite{Bush, Gao}.


The authors thank L. Chen and G. Holland for technical
assistance, and J. A. Liddle and V. Aksyuk for fruitful discussions.
Support from the US NSF (through grants
ECCS-0643178, CBET-0755927, and CMMI-0970071) is acknowledged.


\begin{thebibliography}{1}
\bibitem{shear-mass}The  shear-induced added mass is negligibly small because
$\delta\approx 6\ \mu$m $\ll t_{w}$ when $\Phi_{s}\approx 1$.
\end{thebibliography}

 \end{document}